# Macroscopic single-phase monolayer borophene on arbitrary substrates

*Borna Radatović,\* Valentino Jadriško, Sherif Kamal, Marko Kralj, Dino Novko, Nataša Vujičić, Marin Petrović\**

Center of Excellence for Advanced Materials and Sensing Devices, Institute of Physics, Bijenička 46, 10000 Zagreb, Croatia

**ABSTRACT**

A major challenge in the investigation of all 2D materials is the development of synthesis protocols and tools which would enable their large-scale production and effective manipulation. The same holds for borophene, where experiments are still largely limited to *in situ* characterizations of small-area samples. In contrast, our work is based on millimeter-sized borophene sheets, synthesized on an Ir(111) surface in ultra-high vacuum. Besides high-quality macroscopic synthesis, as confirmed by low-energy electron diffraction (LEED) and atomic force microscopy (AFM), we also demonstrate a successful transfer of borophene from Ir to a Si wafer via electrochemical delamination process. Comparative Raman spectroscopy, in combination with the density functional theory (DFT) calculations, proved that borophene's crystal structure has been preserved in the transfer. Our results demonstrate successful growth and manipulation of large-scale, single-layer borophene sheets with minor defects and ambient stability, thus expediting borophene implementation into more complex systems and devices.

1. **INTRODUCTION**

Borophene (Bo), an atomically thin sheet of boron atoms, gained significant attention in the recent years due to its exceptional physical and chemical properties.[1,2] Due to the polymorphic nature of Bo,[3–5] these properties are essentially variable, which opens pathways for customization of this exceptional 2D material. Many polymorphs have similar formation energies,[4] thus enabling their concurrent experimental realization, but at the same time hindering fabrication of single-polymorph samples. For epitaxial Bo samples, the presence of a substrate can additionally push delicate energetic balance towards the emergence of new polymorphs.[6]

Use of mechanical exfoliation to obtain large high-quality Bo samples is greatly inhibited by the lack of layered bulk boron crystals suitable for this technique. Therefore, Bo research has been notably accelerated by a development of different synthesis methods that enabled fabrication of samples suitable for experimentation. These methods include molecular beam epitaxy (MBE) from solid boron rods,[7–9] sonochemical exfoliation from boron powders[10,11] and chemical vapor deposition (CVD) from gaseous diborane precursor.[12,13] Synthesis efforts are often aimed at production of large-area Bo, pushing Bo research closer to applications in, e.g., supercapacitors, flexible memory devices, photodetectors and batteries.[11,14–16] For example, full monolayer Bo coverage has been reached on Ag(111),[17] Cu(111),[9] Cu(100)[18] and Ir(111).[13,19] Moreover, the monolayer limit has been surpassed by the synthesis of a full Bo bilayer on Ru(0001)[20] and Cu(111).[21] In all these cases, Bo samples were restricted in size only by the size of their substrates.

In parallel to high-quality synthesis, an additional challenge for application development and device fabrication is to manipulate and transfer Bo samples to arbitrary target substrates.[1,22] Only a small number of studies have addressed this crucial issue up to now. Ranjan et al. found that Bo obtained via sonochemical exfoliation in various solvents, followed by centrifugation and placement on different substrates, can provide Bo samples of different polymorphs, thicknesses and particle sizes.[10] Mazaheri et al. performed transfer of CVD-grown Bo from aluminum substrate to target Si substrates.[12] The transfer method consisted of poly(methyl methacrylate) (PMMA) spin coating and treatment of the sample in several liquid substances. This growth-transfer combination also provided many isolated Bo sheets of different sizes, layer count and polymorphs. Chahal and co-workers managed to extract single-layer Bo sheets of dominantly only one polymorph from crystalline boron pieces and place them on different insulating substrates.[23] However, lateral dimensions of the transferred monolayer flakes were in the sub-micrometer range.

Therefore, it remains a challenge to (i) fabricate extended, high-quality Bo samples, with uniform thickness and homogeneous structure, and to (ii) transfer them to an arbitrary substrate with a minimal structural impairment. Here we address these challenges by demonstrating the transfer of macroscopic single-layer Bo sheets from the initial Ir(111) growth substrate to the target Si wafer. Our results prove that deterministic manipulation of Bo layers is possible despite their inherent chemical and mechanical instability, thus expediting Bo research and utilization.

## 2. EXPERIMENTAL METHODS

### 2.1 Synthesis in UHV

Bo samples have been synthesized in an ultra-high vacuum (UHV) system on an Ir(111) substrate (Mateck). Prior to Bo growth, the Ir single-crystal surface was cleaned by several cycles of Ar$^+$ sputtering at 1.5 keV followed by heating in oxygen at 850 °C and annealing at 1200 °C. For Bo synthesis, we employed segregation-enhanced epitaxy method,[19] during which the Ir surface was repeatedly exposed to borazine (B$_3$H$_6$N$_3$, Katchem) vapors at a pressure of 3×10$^{-8}$ mbar and temperature of 1200 °C for 480 s. Each exposure cycle was followed by sample cooling to room temperature, where cooling to 850 °C was carried out at a moderate rate (2 to 3 °C/s) in order to avoid sample quenching and to ensure maximum segregation of boron to the surface. Further cooling preceded at a rate of approximately 10 °C/s. *In situ* LEED was used to inspect Bo crystallography and coverage.

### 2.2 Atomic force microscopy

Atomic force microscopy (AFM) images of Bo were recorded with a JPK Nanowizard Ultra Speed AFM in ambient conditions. Non-contact AC (tapping) mode was used for data acquisition with a setpoint of around ~55%. Bruker TESP-V2 silicon tips with a nominal spring constant of 37 N/m, a tip radius of 7 nm and a resonant frequency of 320 kHZ were used. Images were processed with JPK Data Processing software and WSxM software.[24]

### 2.3 Scanning electron microscopy

Scanning electron microscopy (SEM) characterization of the transferred Bo was performed by in a Tescan VEGA3 microscope with a tungsten cathode. Imaging was performed at a working distance of 10 mm and with 5 kV accelerating voltage. All images were acquired with a secondary electron (SE) detector.

### 2.4 Optical microscopy

Optical images were taken with a home-built microscope in backscattered configuration using Quartz Tungsten-Halogen lamp as white light source, in Koehler illumination configuration with 5× magnification.

**2.5 Raman spectroscopy**

Raman spectra were taken by a home-built micro-Raman system based on confocal microscope in backscattered configuration with a laser excitation of 532 nm (spot size ~0.86 μm, maximum laser power of 1500 μW). The incident laser was focused by a 50× infinity corrected objective (NA = 0.75). The backscattered optical signal was guided through the cage system consisting of a set of Bragg filters after which the signal was coupled to 50 μm core fiber which serves as a confocal detection pinhole. Backscattered light is analyzed in 50 cm long spectrometer (Andor-500i-B1) equipped with three different diffraction gratings (150, 300, 1800 l/mm) and cooled EM CCD detector (Newton 971).

**2.6 Transfer process**

Electrochemical delamination method was used for removal of Bo from Ir crystal after which it was placed on a Si wafer. The method utilizes powder of tetra-n-octylammonium bromide (TOABr) mixed with acetonitrile (also known as methyl cyanide, MECN) for intercalation of large TOA$^+$ ions between Bo monolayer and the Ir crystal. Then, Bo was coated with a layer of PMMA and delaminated from Ir in a 1 M NaOH solution. Detailed description of Bo transfer will be given in Subsection 3.2.

**2.7 Theoretical calculations**

The ground-state density-functional-theory (DFT) calculations were done by means of the Quantum Espresso package[25] with a plane-wave cutoff energy of 70 Ry. Optimized norm-conserving Vanderbilt pseudopotentials were used with the PBE exchange-correlation functional.[26] A freestanding (6×2) hexagonal structure with 5 vacancies and unit cell parameter of a = 5.432 Å was taken as a most stable phase of boron on Ir(111) surface.[27] Two adjacent Bo layers are separated by 30 Å vacuum space in the supercell approach. Momentum space was sampled with a 4×12×1 Monkhorst-Pack grid (with Gaussian smearing of 0.01 Ry). Phonon frequencies at the center of the Brillouin zone were obtained by means of density functional perturbation theory.[28] Final calculated Raman spectra has been generated by applying Lorentzians (with the broadening of 11 cm$^{-1}$ = 1.4 meV) to this set of discrete phonon modes.

## 3. RESULTS AND DISCUSSION

### 3.1 UHV synthesis and pre-transfer characterization

Repetitive cycles of borazine exposure at 1200 °C and subsequent sample cooling were employed to synthesize Bo on Ir(111) [Bo/Ir(111)].[19] During borazine exposure at high temperature, B atoms dissolve in the subsurface regions of Ir. As the sample is cooled after borazine exposure, the solubility of B in Ir reduces and B atoms segregate to the surface of Ir crystal and self-assemble into borophene. The process is schematically illustrated in Figure 1. With each cycle, the amount of dissolved B atoms available for segregation to the Ir surface increases, and Bo coverage ($\Theta_{Bo}$) becomes larger accordingly. Therefore, cycle repetition ensures formation of a full Bo layer after the final B segregation step.

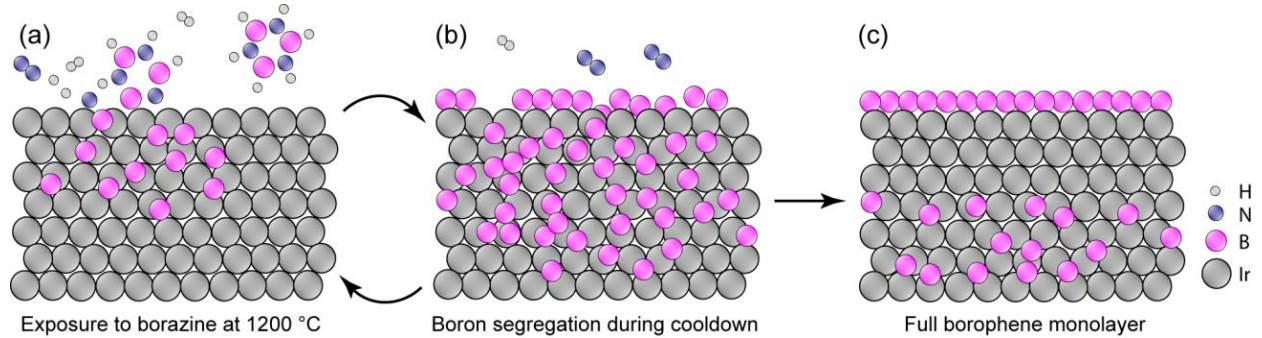

**Figure 1.** Schematic of segregation-enhanced epitaxy of Bo on Ir(111). (a) Borazine precursor is dosed repeatedly onto the hot Ir surface, resulting in boron dissolution into the bulk. (b) Sample cooldown follows each dosing step, during which B atoms segregate to the surface and self-

assemble into a Bo mesh. (c) After finishing a set of dosing/cooling cycles, a full Bo monolayer is formed on the surface.

An increase of $\Theta_{Bo}$ as a function of the number of employed growth cycles ($N_c$) was tracked by low energy electron diffraction (LEED). Figure 2 shows LEED pattern line profiles across the first order Ir diffraction spot and its first neighboring Bo spot at several $N_c$. For $N_c < 4$, Bo spots could not be detected with our LEED apparatus, indicating rather low $\Theta_{Bo}$. As $N_c$ increases, the intensity of the Bo spot rises continuously, indicating significant increase of $\Theta_{Bo}$. At the same time, intensity of the Ir spot diminishes, signifying the formation of an overlayer on top of the Ir surface. An increase of $N_c$ beyond $N_c = 12$ did not yield an additional increase of Bo or decrease of Ir diffraction spot intensities, indicating full Bo layer completion. A typical LEED pattern of 1 ML Bo/Ir(111) is shown in the Figure 2 inset, displaying a characteristic sharp (6×2) pattern with three rotational domains.[19,29] The $\chi_6$ polymorph has been assigned to such a diffraction pattern and the same polymorph can also be fabricated on Ir(111) by employing MBE[27] or conventional CVD technique.[13]

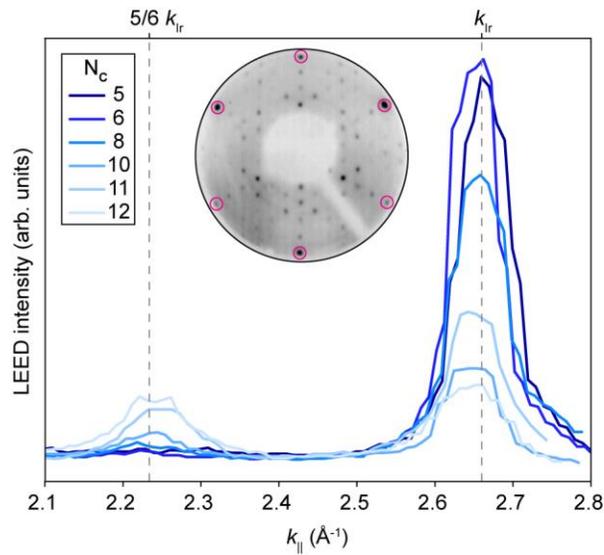

**Figure 2**. A sequence of Ir and Bo diffraction spot profiles extracted from LEED images recorded after the noted number of dosing/cooling cycles ($N_c$). A decrease of Ir spot intensity and an increase of Bo spot intensity as a function of $N_c$ is visible. Inverted LEED image of a monolayer Bo/Ir(111) recorded at 35 eV, with the Ir spots encircled in magenta, is shown in the inset. Superposition of three 120°-rotated (6×2) patterns corresponds to the Bo layer.

After initial *in-situ* confirmation of Bo layer formation, its structural quality and coverage with LEED, samples were taken out of UHV and characterized *ex-situ* with AFM to determine sample morphology and stability under ambient conditions. AFM measurements confirm that essentially the entire surface is covered with a monolayer of Bo, in agreement with the study by Omambac et al. where similar synthesis approach has been employed.[19] This is visible in a large-scale topographic image in Figure 3(a), where Bo spans over many terraces. The corresponding phase image, shown in Figure 3(d), confirms full coverage and the absence of other materials on the surface.

However, it should be noted that Bo can oxidize when exposed to air, as has been shown by *ex-situ* XPS measurements of the $\chi_6$ polymorph.[19] Studies of other polymorphs have shown that only a portion of B atoms get oxidized upon exposure to air, and oxidation is found to be much more pronounced at Bo island edges.[7,12,17,23,30] By taking into account that our sample dominantly consists of fully covered Bo regions, void of free edges, we presume that our Bo sample is not necessarily fully oxidized.

Segregation-assisted epitaxial growth of Bo is known to result in repositioning of Ir steps and formation of wide terraces bounded by step bunches.[19] This is also evident from the line profile shown in Figure 3(g) which has been extracted from Figure 3(a), as indicated by an arrow. The profile shows that the height difference between neighboring Bo-covered terraces ranges up to 30

Å. Since the monoatomic step height on the Ir(111) surface is measured to be 2.4 Å,[31] it is easy to calculate that the step bunch in this case contains as much as thirteen Ir monoatomic steps. Additionally, Ir step bunches intersect under different angles, which is not a feature of clean Ir surfaces and is also attributed to Bo growth process. Generally, *ex-situ* AFM characterization indicates that exposure of Bo to air did not induce any notable morphological changes in the material. It is also important to note that no growth-induced wrinkles have been observed for Bo on Ir(111), similar to hexagonal boron nitride (hBN) on Ir(111)[29] and in contrast to graphene on the same substrate.[32,33] This finding signifies a rather weak interaction between Bo and Ir, or negligible mismatch of thermal expansion coefficients of Bo and Ir.

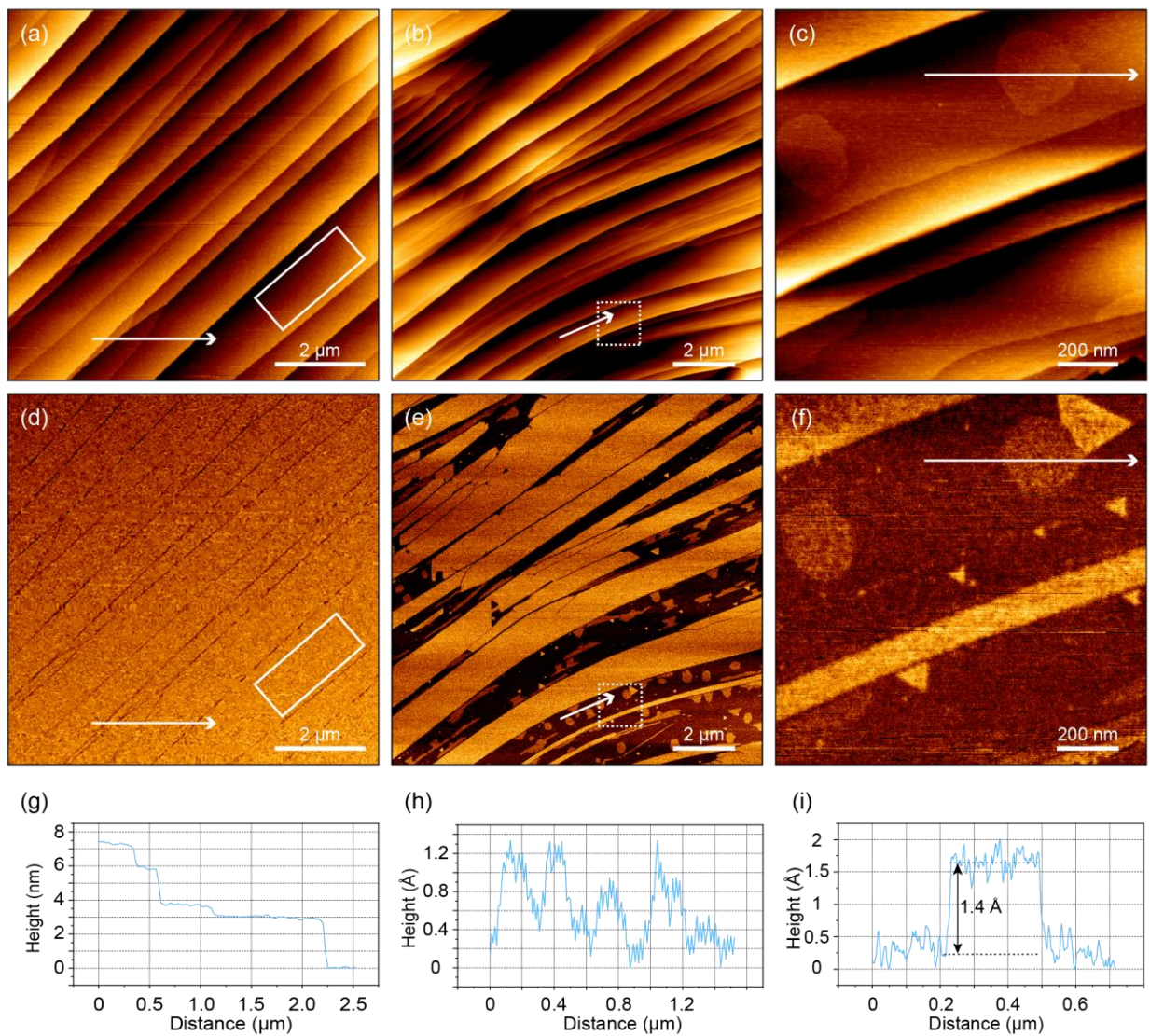

**Figure 3**. AFM imaging of Bo on Ir crystal. (a) Area with a full Bo coverage, white rectangle indicates region inspected for roughness analysis. (b) Area partially covered with Bo sheets and isolated Bo islands. (c) Zoom-in into the area marked in (b) by a white dashed square, with isolated Bo (circular in shape) and residual hBN islands (triangular in shape). (d), (e) and (f) Phase images of (a), (b) and (c), respectively. (g), (h) and (j) Line profiles from (a), (b) and (c), respectively, as indicated by arrows. Image flattening was performed prior to profile extraction.

Systematic AFM scanning across the entire sample surface was performed to find regions with sub-monolayer Bo coverage. Such regions enable topographic AFM characterization which is inaccessible in full-coverage regions, such as Bo height measurement. Those regions are scarce and were dominantly located close to the physical edge of the Ir crystal. One such region is shown in Figure 3(b) and (e) (topography and phase, respectively), where partially covered Ir terraces, sometimes decorated with individual Bo islands, can be seen. These features are obvious in AFM phase image, where dissimilar materials yield different contrasts. The observation of a typical terrace-filling flat material morphology, which has been identified before in LEEM,[19] confirms that the layer covering the Ir surface is indeed Bo.

A zoom-in of region indicated in Figure 3(b) and (e) by a dashed white square is shown in Fig. 3(c) and (f) (topography and phase, respectively). There, individual Bo islands can be clearly resolved, especially in the phase image. A topographic line profile across one such island is shown in Figure 3(i), yielding Bo height of 1.4 Å. However, it is important to note that our measured Bo height varied from 0.9 to 1.5 Å, depending on the tip and the image acquisition parameters used. Such variations are a known property of AFM imaging of 2D materials.[34] For comparison, the reported AFM height of Bo on Cu(111), which is categorized as more weakly bound system,[13,27] is 3 Å.[9]

In addition to the individual Bo islands, regions with $\Theta_{Bo} < 1$ sometimes contain small triangular islands with two distinct, 180°-rotated orientations, as visible in Figure 3(f). These are ascribed to remnants of hBN formation on Ir, which can locally occur due to the nature of the segregation-enhanced Bo growth mechanism: the competition between hBN formation and hBN disintegration accompanied by boron dissolution and resurfacing.[19,29] hBN islands essentially vanish as the Bo coverage increases.

## 3.2 The transfer procedure

To transfer Bo from Ir crystal to an arbitrary substrate, a variant of electrochemical method (often called "the bubbling method") was used, which has been previously employed to successfully transfer graphene and hBN from different metal substrates,[35–37] including Ir(111).[38] In a crucial segment of this method, small hydrogen bubbles form at the metal substrate, gradually expanding across the interface of the metal surface and the 2D material. These bubbles slowly separate delicate material from its synthesis substrate and enable its placement on a target substrate. Our choice of the substrate is a Si wafer with a 280 nm thick $SiO_2$ on top, but essentially any other solid material can be used.

The entire process of Bo transfer is illustrated in Figure 4. Ir crystal with Bo on top [Bo/Ir, step (a)] is immersed into a solution of MECN and TOABr (ratio of 100:1). There, the Ir crystal acts as a cathode at a negative potential and a platinum foil, immersed into the same solution, acts as a positive anode. Upon application of a potential difference of 1.9 V between the electrodes, intercalation of $TOA^+$ ions occurs [step (b)].[39] The purpose of this step is to loosen chemical bond between Bo and Ir substrate. After this first electrochemical step once the sample is taken from a molecular solution and dried, a drop (~100 µL) of PMMA is put on top of the sample and left to dry and solidify [step (c)]. PMMA has a role of a carrier film providing mechanical support to Bo layer in subsequent transfer steps. After that, the second electrochemical step follows, where the PMMA/Bo/Ir sample (the cathode) is immersed into 1M NaOH solution together with the Pt foil (the anode). A ramping potential difference of up to 2.5 V is applied, during which bubbles are formed at the interface of Ir crystal and Bo [step (d)].[38] See also Figure 5(a) for a photograph of the bubbling process.

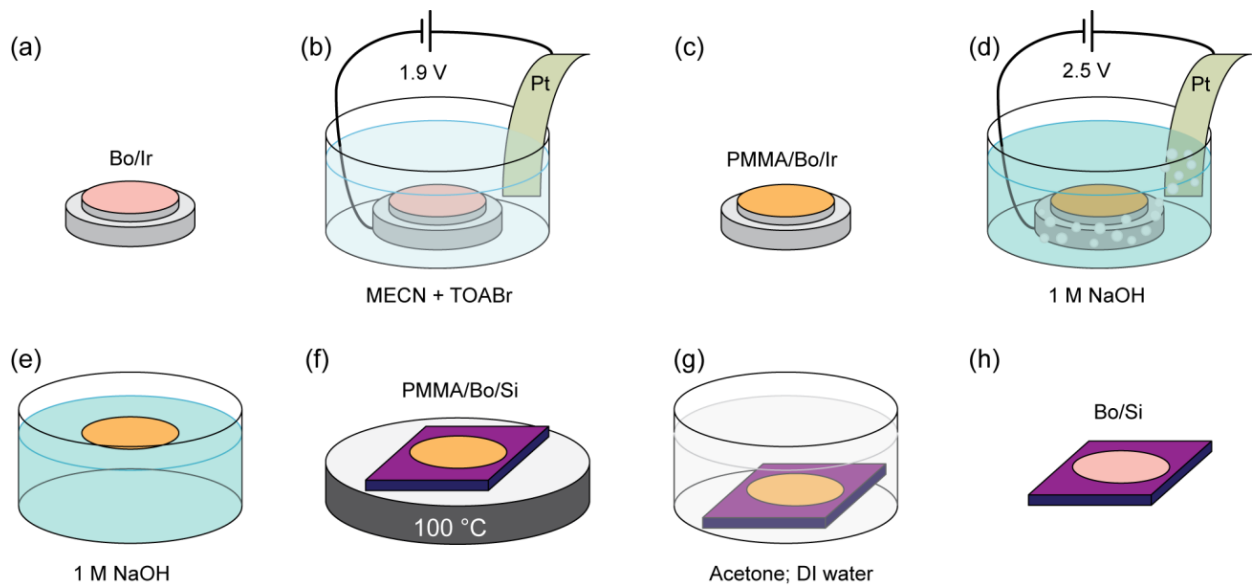

**Figure 4**. Illustration of Bo transfer process. (a) Bo/Ir, (b) Bo/Ir and Pt foil immersed into a solution of MECN and TOABr for the first electrochemical step, (c) PMMA/Bo/Ir, (d) PMMA/Bo/Ir and Pt foil immersed into 1 M NaOH solution for the second electrochemical step, (e) Floating PMMA/Bo, separated from the Ir substrate, (f) PMMA/Bo/Si on a hot plate, (g) PMMA removal with acetone and DI water, (h) Bo/Si.

With the formation of bubbles, PMMA/Bo gradually separates from the Ir crystal (typically within 90 seconds) leaving it floating [step (e)], after which PMMA/Bo is thoroughly rinsed in deionized (DI) water and scooped with a piece of Si wafer (or any other substrate if desired). See Figure 5(b) for a photograph of PMMA/Bo/Si while removed from the Ir crystal. PMMA/Bo/Si is then placed on a hot plate for 30 minutes at 100° C [step (f)], in order to minimize the amount of residual chemicals and reduce transfer-induced wrinkling.[40] After that, PMMA is removed by immersing PMMA/Bo/Si into acetone for ten minutes and subsequent rinsing with DI water [step (g)]. Finally, as a result of the entire process, Bo sample is left on top of an oxidized Si wafer [Bo/Si, step (h)].

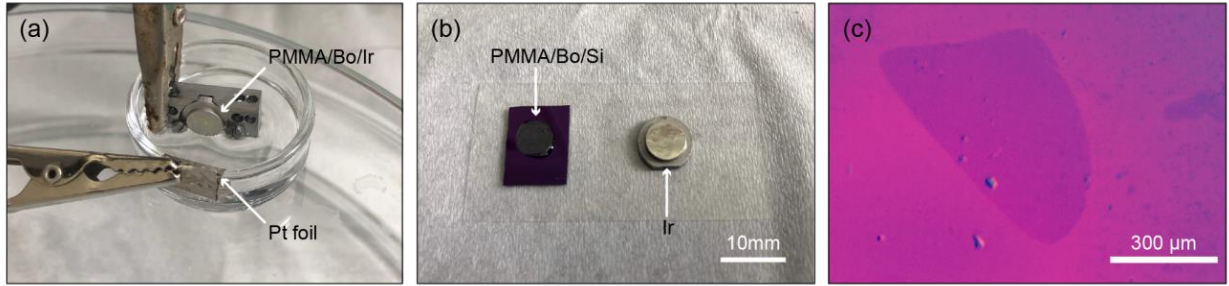

**Figure 5**. Photographs of the transfer process. (a) Formation of bubbles on Ir crystal upon application of a potential difference between Ir and Pt foil. (b) PMMA/Bo placed on a Si wafer (left), after detaching from the Ir crystal (right, still wet from the transfer process). (c) Optical image of the largest transferred fragment of Bo layer.

### 3.3. Post-transfer characterization

Initial inspection of the transferred Bo has been done with an optical microscope. It revealed that the Bo layer suffered partial mechanical damage during the transfer and fragmentation to separated patches occurred. One of the largest fragments is shown in Figure 5(c), exhibiting nearly millimeter lateral dimension. Apparently, as a result of mechanical manipulation and involvement of different chemicals in the transfer process, some parts of Bo have not been transferred or are etched away.

One of the factors which can contribute to fragmentation is the existence of regions on the sample surface where $\Theta_{Bo} < 1$, as shown in Figure 3. Even though these regions are scarce, they represent initial imperfections which can promote etching or tearing of the Bo layer when subjected to a sequence of transfer steps. In addition, since different chemical species readily bind to epitaxial Bo,[7,17,41] involvement of chemicals used in the transfer procedure is likely to promote etching of the Bo layer. Such etching processes might be additionally enhanced due to the confined reaction space at the Bo-Ir interface.[42]

Further characterization of post-transferred Bo has been conducted with AFM and SEM. The flake shown in Figure 5(c) is also shown in SEM topograph in Figure 6(a), where additional details can be resolved. The entire flake appears homogeneous, apart from a few larger cracks and holes (see yellow and cyan arrows, respectively). A zoom-in into the central region of the flake is shown in Figure 6(b), where additional smaller holes can be resolved. These holes appear to be distributed along a series of almost parallel lines that are several micrometers apart. The arrangement of these curvy lines and their direction is strikingly reminiscent of the distribution of Ir step-bunches prior to Bo transfer (cf. Figure 3 and Ref.[19]). Therefore, we suspect that holes in the transferred Bo dominantly appear on locations where Bo/Ir exhibited structural imperfections and possibly different binding - in our case related to Ir step bunches formed during growth. It is plausible that Bo located at the Ir step bunches is strained and chemically more reactive, and therefore more easily damaged during electrochemical treatment and mechanical manipulation. In other words, most of the defects in our transferred Bo stem from inhomogeneity of the synthesis substrate, rather than from the transfer procedure itself. In a similar way, substrate step-edges were found to imprint onto graphene transferred from Cu foils.[35]

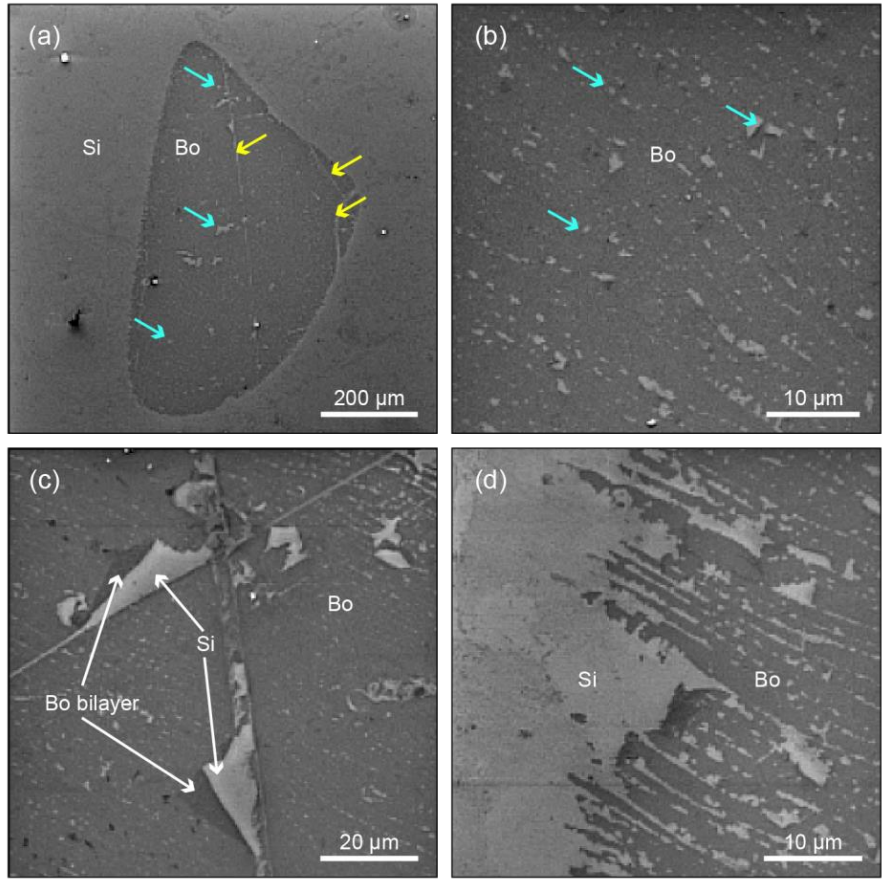

**Figure 6.** SEM images of the transferred Bo. (a) Largest transferred Bo fragment. (b) Magnified view of the central region. (c) Magnified view displaying self-overlapped segments. (d) Magnified view of the edge.

AFM image in Figure 7(a) provides even better resolution of transferred Bo layer. Overall, the sample is flat and free of large-scale defects and thickness inhomogeneities, which indicates that our transfer method did not leave any notable mesoscopic PMMA contamination on the Bo layer. In addition to the holes visible with SEM, quasi-1D defect lines are observed, again mirroring previously discussed Ir step-bunch regions. No conclusive evidence is found of transfer-induced wrinkling of Bo. However, several folds have been identified where Bo layer overlapped onto itself and locally formed a bilayer, as confirmed from the corresponding profile shown in Figure 7(c). Such folds are also clearly distinguishable in SEM [see Figure 6(c)].

At the edge of the transferred Bo flake, the number of holes and cracks in the layer increases, as visible in Figures 6(d) and 7(b). This can be expected, since etching and/or tearing of the Bo layer took place there, as described previously. The same processes most likely resulted in accumulation of excess material in the form of bright protrusions which are visible in AFM topography, extending up to 10 nm in height. Figure 7(b) and the corresponding line profile in Figure 7(d) allow determination of the height of the transferred Bo, which is 3.1 nm. This measured Bo height is more than an order of magnitude larger than the height of Bo on Ir(111).

Besides already discussed variations in AFM height measurements,[34] a dominant factors which could contribute to such an apparently large Bo height increase after the transfer are different substrate and the adsorption of various chemical species the sample surface, both during and after the transfer. During the transfer, depicted in Figure 4, adsorption of atoms and molecules from different solutions to the basal plane of Bo is plausible. In particular, water molecules originating either from exposure to air or from the transfer procedure can form thin layer of different thickness depending on surface affinity for water adsorption (hydrophilic or hydrophobic). Such water layer can significantly influence the measured height of Bo on Ir and on oxide-covered Si, as has been demonstrated for graphene on $SiO_2$.[43]

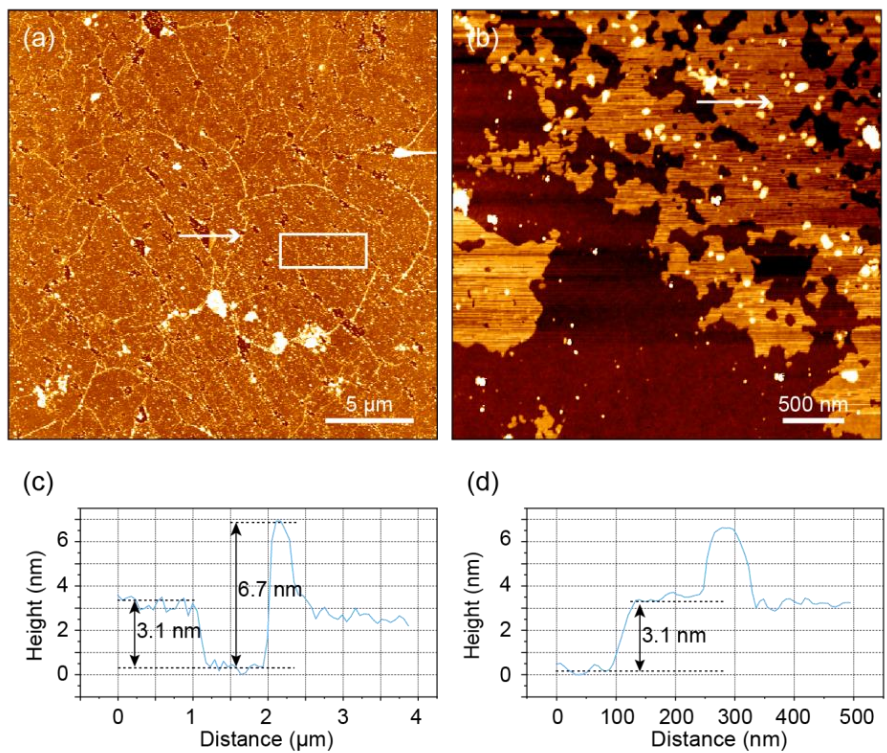

**Figure 7**. AFM imaging of Bo after the transfer to an oxide-covered Si wafer. (a) Topography of the central region of the largest transferred Bo flake, where white rectangle indicates area inspected for roughness analysis. (b) Topography of the edge of the largest transferred Bo flake. (c) and (d) Line profiles extracted from panels (a) and (b) (as marked by arrows), respectively, with heights of the transferred monolayer and bilayer Bo indicated.

Surface roughness (here quantified as root mean square value, RMS roughness) of pre- and post-transfer Bo have been calculated from AFM images, and respective height distribution histograms are shown in Figure 8. RMS roughness of Bo/Ir is 0.73 Å and 15.3 Å of Bo/Si. Roughness of Bo/Ir obtained in our measurement is comparable to roughness of Bo on Cu(111) of 0.43 Å,[9] and signifies a smooth surface of an epitaxially grown 2D material. An increase by a factor of ~20 after the Bo transfer is due to the presence of cracks and fragments in the transferred Bo, as well as the nanoscopic remnants of different chemicals and solvents used in the transfer procedure. Besides post-transfer sample heating [see Figure 4(f)], we propose several other procedures for reduction of Bo roughness in future studies: voltage optimization during electrochemical steps (ramp and final level), substrate flattening (i.e., maximization of terraces widths) and use of other support polymers and polymer removers.

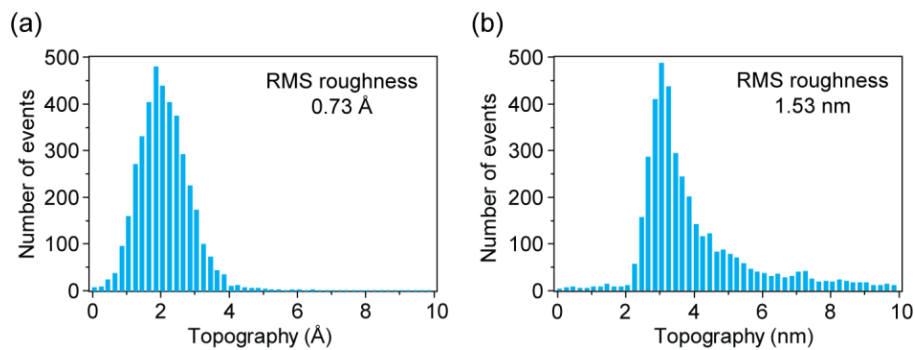

**Figure 8**. Histograms of height distribution of (a) Bo/Ir and (b) Bo/Si, extracted from areas indicated by white rectangles in Figures 3(a) and 7(a), respectively. RMS roughness values are indicated for both cases. Note different units on the x-axes.

A common method to evaluate the success of synthesis and/or transfer of 2D materials is Raman spectroscopy. Raman measurements of Bo sheets can be found in the literature,[10,23,44] but the technique is still not firmly established as a standard Bo characterization tool, as in the case of, e.g., graphene. Ideally, Raman spectroscopy could also be used for identification of different Bo polymorphs. In line with this, we used DFT to calculate vibrational modes and Raman spectra of a freestanding Bo layer ($Bo_{free}$) and we also experimentally measured Raman spectra of Bo/Ir and Bo/Si. The respective data is presented in Figure 9. A large number of Raman peaks can be discerned in all spectra, in accordance with the large unit cell of the Bo $\chi_6$ polymorph which contains 25 B atoms.[27] Many peaks of the Bo/Ir spectra show a good overlap with the peaks of the Bo/Si spectra [Figures 9(b) and (c), see vertical green stripes]. This signifies that the transfer of Bo layer has been successful, preserving the original crystal structure of Bo in the transfer process.

Importantly, no PMMA signal has been detected in the Bo/Si spectra (see Supporting Information), indicating that PMMA residuals have been removed from the material.

Even though the same Raman peaks can be identified in Bo/Ir and Bo/Si spectra, they are found at slightly different energies (of the order of 10 cm$^{-1}$) and have different relative intensities. Energy shifts are expected due to the alterations in interaction and doping induced by the change of the substrate and the chemical environment (i.e., the presence of adsorbates originating from the transfer process). For example, Raman mode shifts have been identified for organic and oxygen adsorbates on Bo.[45,46] Also, strain release within Bo upon the transfer is a possible candidate for the observed energy shifts.[27,32,47] From the technical side, high laser power needed to detect Raman scattering increases the temperature of the sample, which also changes the Raman peak positions of atomically thin materials depending on the type of the substrate.[48]

Peak intensity decrease can also be caused by strain release during the transfer,[49] or by defects and compositional modifications of Bo during different steps of the transfer,[38,50,51] driven by chemical reactivity of Bo already discussed above. Overall, based on the AFM and Raman data we conclude that the transferred material is as-grown Bo layer, likely decorated with foreign species originating either from air exposure or contact with solvents during the transfer process.

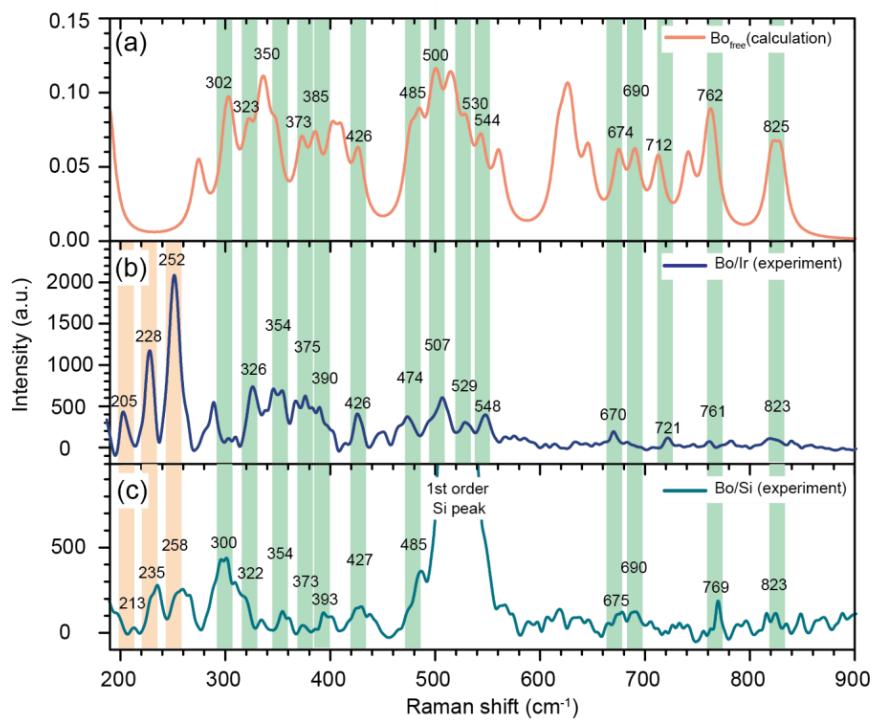

**Figure 9**. Raman characterization of Bo. (a) DFT-calculated Raman spectrum of a freestanding Bo layer (Bo$_{free}$). (b) Measured Raman spectra of Bo/Ir (after background subtraction) and (c) of Bo/Si. Vertical colored stripes are guides to the eye which indicate common Raman modes. Energies (in cm$^{-1}$) of indicated Raman modes are given as numerical values in each panel.

Besides very good correspondence between the Raman spectra of pre- and post-transferred Bo, the majority of experimentally obtained peaks in Figures 9(b) and (c) have their counterparts in the calculated spectra of freestanding Bo sheet (Bo$_{free}$), shown in Figure 9(a). The correspondence between Bo/Ir and Bo$_{free}$ peaks signifies that Bo is weakly bound to the Ir surface. Indeed, the very observation of distinct Raman peaks in the Bo/Ir system is an indication that Bo-Ir interaction is weak, as it has been shown that the Raman spectra of 2D materials can be quenched if the binding to the metal substrate is strong.[38,52]

Despite overall good agreement between our experiments and DFT calculations, some discrepancies can also be found. Experimental low-energy Raman modes (peaks marked by orange vertical stripes in Figure 9) are not reproduced well in DFT calculations of Bo$_{free}$, which has also been seen for the $\chi_3$ and $\beta_{12}$ Bo polymorphs on Ag(111).[44] When bound to the substrate, new Raman modes may become activated in Bo due to different factors such as reduction of symmetry[53–55] or interaction with the substrate.[56] The energy of these so-called Raman interaction modes is substrate-dependent,[47,56–58] which is in line with the shifts of low-energy peaks in Bo/Ir and Bo/Si. The energy positions of other Raman modes are expected to be substrate-dependent to some extent. An obvious example of this is the $B_g$ mode at ~300 cm$^{-1}$ shown schematically in Figure 10(a), exhibiting out-of-plane motion of B atoms. Besides the 300 cm$^{-1}$ mode, Fig. 10 illustrates corresponding phonon modes of Bo which were identified in measured Raman spectra on both substrates.

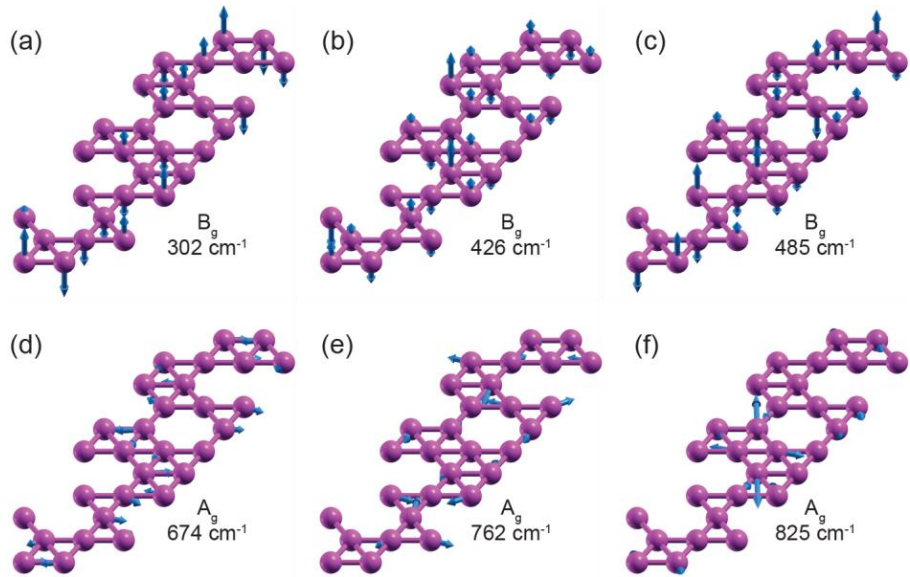

**Figure 10.** Perspective view of six calculated Raman active modes for the freestanding Bo $\chi_6$ polymorph. (a-c) $B_g$ symmetry modes and (d-f) $A_g$ symmetry modes with indicated energies for each mode. The blue arrows and their lengths denote the directions and relative amplitudes of atomic vibrations, respectively.

It should be noted that on the high-energy part of the spectra, Raman peaks of Bo/Ir exhibit reduced intensity and lager widths. Atomic motions of the two representative modes, at ~760 cm$^{-1}$ and ~820 cm$^{-1}$ and both of Ag symmetry, are schematically shown in Figure 10(e) and (f) respectively, displaying clear in-plane motion of B atoms. The absence of pronounced Raman peaks associated with these modes can be explained by enhanced interaction between B-B stretching vibration modes with the boron σ electronic states when metal substrate is present.[59] Such signal attenuation can also lead to apparent merging of closely-spaced Raman peaks, as visible for the ~820 cm$^{-1}$ modes in Figure 9.

## 4. CONCLUSION

In this study, we have linked and successfully executed two crucial steps of Bo research and utilization: large-scale growth of high-quality monolayer samples followed by deterministic transfer of those Bo samples to a targeted substrate. High sample quality is achieved by conducting Bo synthesis on a metallic Ir substrate in controllable UHV conditions. Uniformity and integrity of as-grown Bo monolayer was confirmed by *ex-situ* AFM analysis, disclosing full Bo coverage as well as stability at ambient conditions, void of any notable structural changes. Electrochemical transfer of Bo to SiO$_2$/Si wafer was confirmed by AFM and SEM imaging which showed that macroscopic transfer of Bo is possible. Post-transfer Bo displayed mechanical defects in the form of cracks and holes which were largely inherited from the characteristic imperfections of the synthesis substrate. Further proof of a successful Bo transfer was gained from Raman spectroscopy, where very good overlap of the Raman peaks before and after the transfer confirms

Bo preservation in its original, as-grown crystal structure. Minor quantitative differences in the Raman spectra can be explained by chemical modification of the sample during the transfer, change of the substrate and strain release within the Bo layer. Calculated phonon energies based on DFT calculations of a freestanding Bo $\chi_6$ polymorph are in good agreement with both experimental Raman spectra which indicates the absence of strong binding of Bo to the examined substrates.

Our results represent a major step towards further utilization of Bo in promising applications, e.g., energy storage, optoelectronic and flexible electronic devices, where physical manipulation of 2D boron sheets will be essential.

## ASSOCIATED CONTENT

### Supporting Information

Comparison of experimentally measured Raman spectra of Bo/Si with the measured Raman spectra of PMMA (PDF).

## AUTHOR INFORMATION

### Corresponding Author

Borna Radatović* – bradatovic@ifs.hr

Marin Petrović* – mpetrovic@ifs.hr

### Notes

The authors declare no competing financial interest.

## ACKNOWLEDGMENT

This work was supported by the Croatian Science Foundation, Grants No. UIP-2020-02-1732, No. UIP-2019-04-6869 and No. UIP-2017-05-3869, and by the Center of Excellence for Advanced Materials and Sensing Devices, ERDF Grant No. KK.01.1.1.01.0001.

**REFERENCES**

(1) Zhang, Z.; Penev, E. S.; Yakobson, B. I. Two-Dimensional Boron: Structures, Properties and Applications. *Chem. Soc. Rev.* **2017**, *46* (22), 6746–6763. https://doi.org/10.1039/c7cs00261k.

(2) Li, D.; Gao, J.; Cheng, P.; He, J.; Yin, Y.; Hu, Y.; Chen, L.; Cheng, Y.; Zhao, J. 2D Boron Sheets: Structure, Growth, and Electronic and Thermal Transport Properties. *Adv. Funct. Mater.* **2020**, *30* (8), 1904349. https://doi.org/10.1002/adfm.201904349.

(3) Tang, H.; Ismail-Beigi, S. Novel Precursors for Boron Nanotubes: The Competition of Two-Center and Three-Center Bonding in Boron Sheets. *Phys. Rev. Lett.* **2007**, *99* (11), 115501. https://doi.org/10.1103/PhysRevLett.99.115501.

(4) Penev, E. S.; Bhowmick, S.; Sadrzadeh, A.; Yakobson, B. I. Polymorphism of Two-Dimensional Boron. *Nano Lett.* **2012**, *12* (5), 2441–2445. https://doi.org/10.1021/nl3004754.

(5) Wu, X.; Dai, J.; Zhao, Y.; Zhuo, Z.; Yang, J.; Zeng, X. C. Two-Dimensional Boron Monolayer Sheets. *ACS Nano* **2012**, *6* (8), 7443–7453. https://doi.org/10.1021/nn302696v.

(6) Zhang, Z.; Yang, Y.; Gao, G.; Yakobson, B. I. Two-Dimensional Boron Monolayers Mediated by Metal Substrates. *Angew. Chemie Int. Ed.* **2015**, *54* (44), 13022–13026. https://doi.org/10.1002/anie.201505425.

(7) Mannix, A. J.; Zhou, X. F.; Kiraly, B.; Wood, J. D.; Alducin, D.; Myers, B. D.; Liu, X.; Fisher, B. L.; Santiago, U.; Guest, J. R.; Yacaman, M. J.; Ponce, A.; Oganov, A. R.; Hersam, M. C.; Guisinger, N. P. Synthesis of Borophenes: Anisotropic, Two-Dimensional Boron Polymorphs. *Science (80-. ).* **2015**, *350* (6267), 1513–1516. https://doi.org/10.1126/science.aad1080.

(8) Kiraly, B.; Liu, X.; Wang, L.; Zhang, Z.; Mannix, A. J.; Fisher, B. L.; Yakobson, B. I.; Hersam, M. C.; Guisinger, N. P. Borophene Synthesis on Au(111). *ACS Nano* **2019**, *13* (4), 3816–3822. https://doi.org/10.1021/acsnano.8b09339.

(9) Wu, R.; Drozdov, I. K.; Eltinge, S.; Zahl, P.; Ismail-Beigi, S.; Božović, I.; Gozar, A. Large-Area Single-Crystal Sheets of Borophene on Cu(111) Surfaces. *Nat. Nanotechnol.* **2019**, *14* (1), 44–49. https://doi.org/10.1038/s41565-018-0317-6.

(10) Ranjan, P.; Sahu, T. K.; Bhushan, R.; Yamijala, S. S. R. K. C.; Late, D. J.; Kumar, P.; Vinu, A. Freestanding Borophene and Its Hybrids. *Adv. Mater.* **2019**, *31* (27), 1900353. https://doi.org/10.1002/adma.201900353.

(11) Lin, H.; Shi, H.; Wang, Z.; Mu, Y.; Li, S.; Zhao, J.; Guo, J.; Yang, B.; Wu, Z.-S.; Liu, F. Scalable Production of Freestanding Few-Layer $β$ 12 -Borophene Single Crystalline Sheets as Efficient Electrocatalysts for Lithium–Sulfur Batteries. *ACS Nano* **2021**, *15* (11), 17327–17336. https://doi.org/10.1021/acsnano.1c04961.

(12) Mazaheri, A.; Javadi, M.; Abdi, Y. Chemical Vapor Deposition of Two-Dimensional Boron Sheets by Thermal Decomposition of Diborane. *ACS Appl. Mater. Interfaces* **2021**, *13* (7), 8844–8850. https://doi.org/10.1021/acsami.0c22580.

(13) Cuxart, M. G.; Seufert, K.; Chesnyak, V.; Waqas, W. A.; Robert, A.; Bocquet, M.; Duesberg, G. S.; Sachdev, H.; Auwärter, W. Borophenes Made Easy. *Sci. Adv.* **2021**, *7* (45), 1–8. https://doi.org/10.1126/sciadv.abk1490.

(14) Li, H.; Jing, L.; Liu, W.; Lin, J.; Tay, R. Y.; Tsang, S. H.; Teo, E. H. T. Scalable Production of Few-Layer Boron Sheets by Liquid-Phase Exfoliation and Their Superior Supercapacitive Performance. *ACS Nano* **2018**, *12* (2), 1262–1272. https://doi.org/10.1021/acsnano.7b07444.

(15) Hou, C.; Tai, G.; Hao, J.; Sheng, L.; Liu, B.; Wu, Z. Ultrastable Crystalline Semiconducting Hydrogenated Borophene. *Angew. Chemie* **2020**, *132* (27), 10911–10917. https://doi.org/10.1002/ange.202001045.

(16) Wu, Z.; Tai, G.; Liu, R.; Hou, C.; Shao, W.; Liang, X.; Wu, Z. Van Der Waals Epitaxial Growth of Borophene on a Mica Substrate toward a High-Performance Photodetector. *ACS Appl. Mater. Interfaces* **2021**, *13* (27), 31808–31815. https://doi.org/10.1021/acsami.1c03146.

(17) Feng, B.; Zhang, J.; Zhong, Q.; Li, W.; Li, S.; Li, H.; Cheng, P.; Meng, S.; Chen, L.; Wu, K. Experimental Realization of Two-Dimensional Boron Sheets. *Nat. Chem.* **2016**, *8* (6), 563–568. https://doi.org/10.1038/nchem.2491.

(18) Wu, R.; Eltinge, S.; Drozdov, I. K.; Gozar, A.; Zahl, P.; Sadowski, J. T.; Ismail-Beigi, S.; Božović, I. Micrometre-Scale Single-Crystalline Borophene on a Square-Lattice Cu(100) Surface. *Nat. Chem.* **2022**, *14* (4), 377–383. https://doi.org/10.1038/s41557-021-00879-9.

(19) Omambac, K. M.; Petrović, M.; Bampoulis, P.; Brand, C.; Kriegel, M. A.; Dreher, P.;

Janoschka, D.; Hagemann, U.; Hartmann, N.; Valerius, P.; Michely, T.; Meyer Zu Heringdorf, F. J.; Horn-Von Hoegen, M. Segregation-Enhanced Epitaxy of Borophene on Ir(111) by Thermal Decomposition of Borazine. *ACS Nano* **2021**, *15* (4), 7421–7429. https://doi.org/10.1021/acsnano.1c00819.

(20) Sutter, P.; Sutter, E. Large-Scale Layer-by-Layer Synthesis of Borophene on Ru(0001). *Chem. Mater.* **2021**, *33* (22), 8838–8843. https://doi.org/10.1021/acs.chemmater.1c03061.

(21) Chen, C.; Lv, H.; Zhang, P.; Zhuo, Z.; Wang, Y.; Ma, C.; Li, W.; Wang, X.; Feng, B.; Cheng, P.; Wu, X.; Wu, K.; Chen, L. Synthesis of Bilayer Borophene. *Nat. Chem.* **2022**, *14* (1), 25–31. https://doi.org/10.1038/s41557-021-00813-z.

(22) Chand, H.; Kumar, A.; Krishnan, V. Borophene and Boron-Based Nanosheets: Recent Advances in Synthesis Strategies and Applications in the Field of Environment and Energy. *Adv. Mater. Interfaces* **2021**, *8* (15), 2100045. https://doi.org/10.1002/ADMI.202100045.

(23) Chahal, S.; Ranjan, P.; Motlag, M.; Yamijala, S. S. R. K. C.; Late, D. J.; Sadki, E. H. S.; Cheng, G. J.; Kumar, P. Borophene via Micromechanical Exfoliation. *Adv. Mater.* **2021**, *33* (34), 2102039. https://doi.org/10.1002/adma.202102039.

(24) Horcas, I.; Fernández, R.; Gómez-Rodríguez, J. M.; Colchero, J.; Gómez-Herrero, J.; Baro, A. M. WSXM : A Software for Scanning Probe Microscopy and a Tool for Nanotechnology. *Rev. Sci. Instrum.* **2007**, *78* (1), 13705. https://doi.org/10.1063/1.2432410.

(25) Giannozzi, P.; Andreussi, O.; Brumme, T.; Bunau, O.; Buongiorno Nardelli, M.; Calandra, M.; Car, R.; Cavazzoni, C.; Ceresoli, D.; Cococcioni, M.; Colonna, N.; Carnimeo, I.; Dal Corso, A.; De Gironcoli, S.; Delugas, P.; Distasio, R. A.; Ferretti, A.; Floris, A.; Fratesi, G.;

Fugallo, G.; Gebauer, R.; Gerstmann, U.; Giustino, F.; Gorni, T.; Jia, J.; Kawamura, M.; Ko, H. Y.; Kokalj, A.; Kücükbenli, E.; Lazzeri, M.; Marsili, M.; Marzari, N.; Mauri, F.; Nguyen, N. L.; Nguyen, H. V.; Otero-De-La-Roza, A.; Paulatto, L.; Poncé, S.; Rocca, D.; Sabatini, R.; Santra, B.; Schlipf, M.; Seitsonen, A. P.; Smogunov, A.; Timrov, I.; Thonhauser, T.; Umari, P.; Vast, N.; Wu, X.; Baroni, S. Advanced Capabilities for Materials Modelling with Quantum ESPRESSO. *J. Phys. Condens. Matter* **2017**, *29* (46), 465901. https://doi.org/10.1088/1361-648X/AA8F79.

(26) Hamann, D. R. Optimized Norm-Conserving Vanderbilt Pseudopotentials. *Phys. Rev. B - Condens. Matter Mater. Phys.* **2013**, *88* (8), 085117. https://doi.org/10.1103/PHYSREVB.88.085117/FIGURES/6/MEDIUM.

(27) Vinogradov, N. A.; Lyalin, A.; Taketsugu, T.; Vinogradov, A. S.; Preobrajenski, A. Single-Phase Borophene on Ir(111): Formation, Structure, and Decoupling from the Support. *ACS Nano* **2019**, *13* (12), 14511–14518. https://doi.org/10.1021/acsnano.9b08296.

(28) Baroni, S.; De Gironcoli, S.; Dal Corso, A.; Giannozzi, P. Phonons and Related Crystal Properties from Density-Functional Perturbation Theory. *Rev. Mod. Phys.* **2001**, *73* (2), 515. https://doi.org/10.1103/RevModPhys.73.515.

(29) Petrović, M.; Hagemann, U.; Horn-von Hoegen, M.; Meyer zu Heringdorf, F.-J. Microanalysis of Single-Layer Hexagonal Boron Nitride Islands on Ir(111). *Appl. Surf. Sci.* **2017**, *420*, 504–510. https://doi.org/10.1016/j.apsusc.2017.05.155.

(30) Liu, X.; Rahn, M. S.; Ruan, Q.; Yakobson, B. I.; Hersam, M. C. Probing Borophene Oxidation at the Atomic Scale. *Nanotechnology* **2022**, *33* (23), 235702.

https://doi.org/10.1088/1361-6528/ac56bd.

(31) Wolfschmidt, H.; Baier, C.; Gsell, S.; Fischer, M.; Schreck, M.; Stimming, U. STM, SECPM, AFM and Electrochemistry on Single Crystalline Surfaces. *Materials (Basel).* **2010**, *3* (8), 4196–4213. https://doi.org/10.3390/ma3084196.

(32) N'Diaye, A. T.; van Gastel, R.; Martínez-Galera, A. J.; Coraux, J.; Hattab, H.; Wall, D.; zu Heringdorf, F.-J. M.; Hoegen, M. H.; Gómez-Rodríguez, J. M.; Poelsema, B.; Busse, C.; Michely, T. In Situ Observation of Stress Relaxation in Epitaxial Graphene. *New J. Phys.* **2009**, *11* (11), 113056. https://doi.org/10.1088/1367-2630/11/11/113056.

(33) Petrović, M.; Sadowski, J. T.; Šiber, A.; Kralj, M. Wrinkles of Graphene on Ir(1 1 1): Macroscopic Network Ordering and Internal Multi-Lobed Structure. *Carbon N. Y.* **2015**, *94*, 856–863. https://doi.org/10.1016/j.carbon.2015.07.059.

(34) Shearer, C. J.; Slattery, A. D.; Stapleton, A. J.; Shapter, J. G.; Gibson, C. T. Accurate Thickness Measurement of Graphene. *Nanotechnology* **2016**, *27* (12), 125704. https://doi.org/10.1088/0957-4484/27/12/125704.

(35) Wang, Y.; Zheng, Y.; Xu, X.; Dubuisson, E.; Bao, Q.; Lu, J.; Loh, K. P. Electrochemical Delamination of CVD-Grown Graphene Film: Toward the Recyclable Use of Copper Catalyst. *ACS Nano* **2011**, *5* (12), 9927–9933. https://doi.org/10.1021/nn203700w.

(36) Gao, L.; Ren, W.; Xu, H.; Jin, L.; Wang, Z.; Ma, T.; Ma, L.-P.; Zhang, Z.; Fu, Q.; Peng, L.-M.; Bao, X.; Cheng, H.-M. Repeated Growth and Bubbling Transfer of Graphene with Millimetre-Size Single-Crystal Grains Using Platinum. *Nat. Commun.* **2012**, *3* (1), 699. https://doi.org/10.1038/ncomms1702.

(37) Kim, G.; Jang, A.-R.; Jeong, H. Y.; Lee, Z.; Kang, D. J.; Shin, H. S. Growth of High-Crystalline, Single-Layer Hexagonal Boron Nitride on Recyclable Platinum Foil. *Nano Lett.* **2013**, *13* (4), 1834–1839. https://doi.org/10.1021/nl400559s.

(38) Koefoed, L.; Kongsfelt, M.; Ulstrup, S.; Čabo, A. G.; Cassidy, A.; Whelan, P. R.; Bianchi, M.; Dendzik, M.; Pizzocchero, F.; Jørgensen, B.; Bøggild, P.; Hornekær, L.; Hofmann, P.; Pedersen, S. U.; Daasbjerg, K. Facile Electrochemical Transfer of Large-Area Single Crystal Epitaxial Graphene from Ir(1 1 1). *J. Phys. D. Appl. Phys.* **2015**, *48* (11), 115306. https://doi.org/10.1088/0022-3727/48/11/115306.

(39) Paredes, J. I.; Munuera, J. M. Recent Advances and Energy-Related Applications of High Quality/Chemically Doped Graphenes Obtained by Electrochemical Exfoliation Methods. *J. Mater. Chem. A* **2017**, *5* (16), 7228–7242. https://doi.org/10.1039/C7TA01711A.

(40) Lu, S.; Jia, H.; Gu -, Z.; Ma, L.; Wang, C.; Chu, Y.; -, al; Pan, M.; Yuan, H.; Cao -, Y.; Jain, A.; Bharadwaj, P.; Heeg, S.; Parzefall, M.; Taniguchi, T.; Watanabe, K.; Novotny, L. Minimizing Residues and Strain in 2D Materials Transferred from PDMS. *Nanotechnology* **2018**, *29* (26), 265203. https://doi.org/10.1088/1361-6528/AABD90.

(41) Li, Q.; Kolluru, V. S. C.; Rahn, M. S.; Schwenker, E.; Li, S.; Hennig, R. G.; Darancet, P.; Chan, M. K. Y.; Hersam, M. C. Synthesis of Borophane Polymorphs through Hydrogenation of Borophene. *Science (80-. ).* **2021**, *371* (6534), 1143–1148. https://doi.org/10.1126/science.abg1874.

(42) Fu, Q.; Bao, X. Surface Chemistry and Catalysis Confined under Two-Dimensional Materials. *Chem. Soc. Rev.* **2017**, *46* (7), 1842–1874. https://doi.org/10.1039/C6CS00424E.

(43) Nemes-Incze, P.; Osváth, Z.; Kamarás, K.; Biró, L. P. Anomalies in Thickness Measurements of Graphene and Few Layer Graphite Crystals by Tapping Mode Atomic Force Microscopy. *Carbon N. Y.* **2008**, *46* (11), 1435–1442. https://doi.org/10.1016/j.carbon.2008.06.022.

(44) Sheng, S.; Wu, J. Bin; Cong, X.; Zhong, Q.; Li, W.; Hu, W.; Gou, J.; Cheng, P.; Tan, P. H.; Chen, L.; Wu, K. Raman Spectroscopy of Two-Dimensional Borophene Sheets. *ACS Nano* **2019**, *13* (4), 4133–4139. https://doi.org/10.1021/acsnano.8b08909.

(45) Li, L.; Schultz, J. F.; Mahapatra, S.; Liu, X.; Shaw, C.; Zhang, X.; Hersam, M. C.; Jiang, N. Angstrom-Scale Spectroscopic Visualization of Interfacial Interactions in an Organic/Borophene Vertical Heterostructure. *J. Am. Chem. Soc.* **2021**, *143* (38), 15624–15634. https://doi.org/10.1021/jacs.1c04380.

(46) Li, L.; Schultz, J. F.; Mahapatra, S.; Lu, Z.; Zhang, X.; Jiang, N. Chemically Identifying Single Adatoms with Single-Bond Sensitivity during Oxidation Reactions of Borophene. *Nat. Commun.* **2022**, *13* (1), 1796. https://doi.org/10.1038/s41467-022-29445-8.

(47) Velický, M.; Rodriguez, A.; Bouša, M.; Krayev, A. V.; Vondráček, M.; Honolka, J.; Ahmadi, M.; Donnelly, G. E.; Huang, F.; Abruña, H. D.; Novoselov, K. S.; Frank, O. Strain and Charge Doping Fingerprints of the Strong Interaction between Monolayer $MoS_2$ and Gold. *J. Phys. Chem. Lett.* **2020**, *11* (15), 6112–6118. https://doi.org/10.1021/acs.jpclett.0c01287.

(48) Zhang, X.; Qiao, X. F.; Shi, W.; Wu, J. Bin; Jiang, D. S.; Tan, P. H. Phonon and Raman Scattering of Two-Dimensional Transition Metal Dichalcogenides from Monolayer,

Multilayer to Bulk Material. *Chem. Soc. Rev.* **2015**, *44* (9), 2757–2785. https://doi.org/10.1039/c4cs00282b.

(49) Zhang, Y.; Guo, H.; Sun, W.; Sun, H.; Ali, S.; Zhang, Z.; Saito, R.; Yang, T. Scaling Law for Strain Dependence of Raman Spectra in Transition-Metal Dichalcogenides. *J. Raman Spectrosc.* **2020**, *51* (8), 1353–1361. https://doi.org/10.1002/jrs.5908.

(50) Ciuk, T.; Pasternak, I.; Krajewska, A.; Sobieski, J.; Caban, P.; Szmidt, J.; Strupinski, W. Properties of Chemical Vapor Deposition Graphene Transferred by High-Speed Electrochemical Delamination. *J. Phys. Chem. C* **2013**, *117* (40), 20833–20837. https://doi.org/10.1021/jp4032139.

(51) Wong, C. H. A.; Pumera, M. Electrochemical Delamination and Chemical Etching of Chemical Vapor Deposition Graphene: Contrasting Properties. *J. Phys. Chem. C* **2016**, *120* (8), 4682–4690. https://doi.org/10.1021/acs.jpcc.6b00329.

(52) Starodub, E.; Bostwick, A.; Moreschini, L.; Nie, S.; Gabaly, F. El; McCarty, K. F.; Rotenberg, E. In-Plane Orientation Effects on the Electronic Structure, Stability, and Raman Scattering of Monolayer Graphene on Ir(111). *Phys. Rev. B* **2011**, *83* (12), 125428. https://doi.org/10.1103/PhysRevB.83.125428.

(53) Chow, C. M.; Yu, H.; Jones, A. M.; Yan, J.; Mandrus, D. G.; Taniguchi, T.; Watanabe, K.; Yao, W.; Xu, X. Unusual Exciton-Phonon Interactions at van Der Waals Engineered Interfaces. *Nano Lett.* **2017**, *17* (2), 1194–1199. https://doi.org/10.1021/acs.nanolett.6b04944.

(54) Amsalem, P.; Giovanelli, L.; Themlin, J. M.; Angot, T. Electronic and Vibrational

Properties at the ZnPc/Ag(110) Interface. *Phys. Rev. B - Condens. Matter Mater. Phys.* **2009**, *79* (23), 235426. https://doi.org/10.1103/PhysRevB.79.235426.

(55) Scheuschner, N.; Gillen, R.; Staiger, M.; Maultzsch, J. Interlayer Resonant Raman Modes in Few-Layer MoS2. *Phys. Rev. B - Condens. Matter Mater. Phys.* **2015**, *91* (23), 235409. https://doi.org/10.1103/PhysRevB.91.235409.

(56) Zhao, Y.; Luo, X.; Zhang, J.; Wu, J.; Bai, X.; Wang, M.; Jia, J.; Peng, H.; Liu, Z.; Quek, S. Y.; Xiong, Q. Interlayer Vibrational Modes in Few-Quintuple-Layer Bi2Te3 and Bi2Se3 Two-Dimensional Crystals: Raman Spectroscopy and First-Principles Studies. *Phys. Rev. B - Condens. Matter Mater. Phys.* **2014**, *90* (24), 245428. https://doi.org/10.1103/PhysRevB.90.245428.

(57) Xiang, Q.; Yue, X.; Wang, Y.; Du, B.; Chen, J.; Zhang, S.; Li, G.; Cong, C.; Yu, T.; Li, Q.; Jin, Y. Unveiling the Origin of Anomalous Low-Frequency Raman Mode in CVD-Grown Monolayer WS2. *Nano Res.* **2021**, *14* (11), 4314–4320. https://doi.org/10.1007/s12274-021-3769-1.

(58) Liang, L.; Zhang, J.; Sumpter, B. G.; Tan, Q. H.; Tan, P. H.; Meunier, V. Low-Frequency Shear and Layer-Breathing Modes in Raman Scattering of Two-Dimensional Materials. *ACS Nano* **2017**, *11* (12), 11777–11802. https://doi.org/10.1021/acsnano.7b06551.

(59) Gao, M.; Yan, X. W.; Wang, J.; Lu, Z. Y.; Xiang, T. Electron-Phonon Coupling in a Honeycomb Borophene Grown on Al(111) Surface. *Phys. Rev. B* **2019**, *100* (2), 024503. https://doi.org/10.1103/PhysRevB.100.024503.